\documentclass[12pt]{article}
\usepackage{psfig}
\newcommand{\beq}{\begin{equation}}
\newcommand{\eeq}{\end{equation}}
\newcommand{\beqn}{\begin{eqnarray}}
\newcommand{\eeqn}{\end{eqnarray}}
\newcommand{\bea}[1]{\beq\begin{array}{#1}}
\newcommand{\eea}{\end{array}\eeq}

\newcommand{\cD}{{\cal D}}

\newcommand{\AP}[3]{{\it Ann. Phys. }{\bf #1} (#2) #3}
\newcommand{\NP}[3]{{\it Nucl. Phys. }{\bf #1} (#2) #3}

\newcommand{\PL}[3]{{\it Phys. Lett. }{\bf #1} (#2) #3}

\newcommand{\MPL}[3]{{\it Mod. Phys. Lett. }{\bf #1} (#2) #3}

\def\U1{ \langle (A_{\mu})^2_{min} \rangle}
\def\u1{  A^2 }
\def\min{$ A^2_{min}$~}
\begin{document}
\date{}
\title{\bf\Large On the Significance of the Quantity $A^2$
\vskip-40mm
\rightline{\small\rm MPI-PhT/00-40}
\vskip 40mm 
}
\author{
F.V.~Gubarev$^{\rm a,b}$, L. Stodolsky$^b$, V.I. Zakharov$^b$ \\
$^{\rm a}$ {\small\it Institute of Theoretical and  Experimental
Physics, B.Cheremushkinskaya 25,}\\
           {\small\it Moscow, 117259, Russia}\\
$^{\rm b}$ {\small\it Max-Planck Institut f\"ur Physik, F\"ohringer
Ring 6, 80805 M\"unchen, Germany}} 

\maketitle
\begin{abstract}\noindent
We consider the gauge potential $A$ and
argue that the minimum   value of the volume integral of $A^2$ (in
Euclidean space) 
may have physical meaning, particularly in connection with the
existence of topological structures. A lattice simulation comparing
compact and non-compact ``photodynamics'' shows a jump in this
quantity at the phase transition, supporting this idea. 
\end{abstract}

\section{Introduction}
 
Physical quantities should be gauge invariant. At first glance
this might seem to imply that only expressions involving the
fields ($E$ and $B$ in electromagnetism) and not the potentials
($A$) should appear in physically  meaningful quantities, and in
fact this is usually true.  However, this logic can  be
misleading. A well known case in point is the loop integral
$\oint {\bf A dx}$. Although only $A$,  and not the fields,
appears explicitly in this construction, it is so devised
that it leads to a gauge invariant and indeed an
interesting object.

We would like to point out the interest of another quantity
constructed  from $A$ itself: the volume integral of $A^2(x)$. One
may come upon this thought when considering
the role of condensates in quantum field theory.
Vacuum condensates  have been a useful way to understand and
characterize the  dynamics of QCD and other field theories. The
most famous example is perhaps the quark condensate:
\begin{equation}
\langle 0|\bar{q}q|0\rangle \neq 0 \,,
\end{equation}
where $q$ stands for light $u$- or $d$-quarks. In the realistic
case of negligibly small quark mass, a nonvanishing value of the
quark condensate signals spontaneous  breaking of chiral symmetry.
 
In the framework of the QCD sum rules \cite{svz} one also used the
concept of the  gluon condensate
\begin{equation}
\langle 0|\alpha_s(G^a_{\mu\nu})^2|0\rangle \neq 0\,.
\label{gc}
\end{equation}
Here the non-vanishing value of the condensate  signifies not the
breaking of a symmetry but
rather the presence of nonperturbative fields in the vacuum.
 
This gluon condensate would appear to be the simplest quantity
characterizing nonperturbative vacuum fields. It has dimension
$\mathrm{d}=4$, leading  one to assume that the leading
non-perturbative corrections in the QCD sum rules  at large
external momentum $Q$ are of order
$\langle 0|\alpha_s(G^a_{\mu\nu})^2|0\rangle /Q^4$. 

Now  there is of course an even simpler  candidate for a
condensate, namely just the square of the vector potential: $\u1$.
This is of dimension $\mathrm{d}=2$. 
However, such expressions  seem  not to  be  allowed since they
appear gauge noninvariant \cite{comment}. That is, one tends to
think that physically meaningful  quantities must 
involve only the fields and not the
potentials, and that an expression like $\u1$, involving only
potentials, could
not be meaningful. However this is not necessarily true, as we
would now like to illustrate on the simple example of
magnetostatics.

\section{Magnetostatics}

 Consider a situation with some
magnetic field ${\bf B}$  present in space. There is a considerable
amount of freedom in the choice
of ${\bf A}$. However since there is a nonzero magnetic field ${\bf
B} ={\nabla}\times  {\bf A}$, we know {\it some} nonzero 
${\bf A}$ must be present; ${\bf A}$
cannot be zero everywhere. Now consider the volume integral of
$A^2(x)$.  It is a positive quantity and  cannot be zero.
  It must then have some
minimum value.
Therefore of all the possible ${\bf A}$ configurations which yield
the
given ${\bf B}$ the one (or the ones)
with the smallest integral of ${\bf A}^2$ has in a sense an
invariant significance.  We would then like
to examine the possible significance of the resulting  quantity,
the
volume
integral of $A^2(x)$ at its minimum value. We will call this \min.

The connection between the ``minimum $A^2$" requirement and a more
familiar gauge condition may be seen as follows. Suppose for a
given field congifuration that $\int {\bf A}^2 d^3x$ is at
its minimum value; then under a gauge transformation it is
stationary. Considering
${\bf A}\rightarrow {\bf A}+{\bf\nabla} \phi$ for infinitesimal
$\phi$  we  have $\int {\bf
A}\,{\nabla}\phi\, d^3x=0$ and  integrating by parts 
\beq
\label{div}
\int \phi\;{\bf \nabla  A} \; d^3 x  ~+~ surface~terms~=~0\,.
\eeq
Since $\phi$ is arbitrary we conclude that, up to the  surface
terms and the question of local minima in $A^2$,
the ``minimum $A^2$" condition is
equivalent to the  familiar gauge condition
\beq
\label{diva}
 {\bf  \nabla  A}~=~0\,.
\eeq
 Not surprisingly the ``minimum $A^2$" requirement is connected
with
that  gauge condition which is invariant, i.e. makes no reference
to any
particular direction.

Furthermore it appears that \min is
sensitive to, or measures in some way, topological features of the
system under consideration. 
This is suggested by the 
 comparison of  two  situations, both  with no magnetic field. Let 
one be
simple
empty space with ${\bf B=0}$, while the other has
a non-trivial topology with the presence of a   tube or string
containing magnetic flux, like a 
``cosmic string''  or a vortex in superconductivity.
In the first case we  have simply no
$A$ and  so \min$=0$. In the second case,  due to the flux $\Phi$
in the
tube   or string
\beq
\oint{\bf A dx}~=~\int {\bf H\cdot }d{\bf s}~\equiv~\Phi\,,
\eeq
and $A$ cannot be zero in the surrounding space even 
though  the magnetic  field  is absent. This example suggests that
\min  can
signal
the presence of non-trivial topological structures.

The logical situation concerning \min  resembles somewhat  that of
the
question of
the energy of a particle in
relativity. The energy of a particle is of course a frame dependent
quantity. However the  minimum energy, which is
the energy in the rest frame, has an invariant meaning, namely the
mass. 
In going to the rest frame of the particle we do make a certain 
choice of  frame,
but nevertheless the mass is an undeniably meaningful
quantity~\cite{sc}. 

Of course the mass also has an explicitly invariant expression,
$m^2= E^2 - P^2$, and  the loop integral $\oint {\bf A} dx$
can, via
Stokes theorem, be expressed in term of the fields.  Analogously,
is there an a 
expression  for  \min directly in terms of the fields? 

 Indeed there is the vector relation~\cite{kcross}
$$
\int {\bf A}^2(x) d^3x =
{1 \over 4\pi}\int d^3x d^3x' 
     {
          [ {\bf\nabla}\times{\bf A}(x)] \cdot
[{\bf\nabla}\times{\bf A}(x')]
     \over 
          |{\bf x}-{\bf x}'|
     }
~+~ {1 \over 4\pi}\int d^3x d^3x'
     {
          [{\bf\nabla}\cdot{\bf A}(x)] [{\bf\nabla}\cdot{\bf
A}(x')]
     \over
          |{\bf x}-{\bf x}'|
     }
$$
\beq \label{vect}
+~ surface~ terms\,.
\eeq
Each of the two terms is positive, hence (up to the surface term
question) we can  minimize the integral
of ${\bf A}^2$ by choosing ${\bf \nabla} \cdot {\bf A}=0$. With
this choice the integral of ${\bf A}^2$
is minimal in accord with our above remarks, and is expressed only
in terms of the magnetic field
${\bf \nabla} \times {\bf A}$:
\beq
\label{mn}
 A^2_{min}~=~
{1 \over 4\pi}\int d^3x d^3x'
     {
          {\bf B}({\bf x})\cdot{\bf B}({\bf x}')
      \over
          |{\bf x}-{\bf x}'|
     }
~+~ surface~terms\,.
\eeq
Thus we can trade, so to speak, apparent locality for explicit
gauge invariance.

 It will be seen that the arguments of this section carry over to
four (or more ) dimensions directly, as long as the metric is
euclidean. For example in four dimensions~Eq[\ref{vect}] becomes
$$
\int A^2(x) d^4x ~=~
{1 \over 2\pi^2}\int d^4x d^4x'
     {
          [F_{\mu\nu}(x)] [F_{\mu\nu}(x')]
     \over
          (x-x')^2
     } 
~+~
{1 \over 2\pi^2}\int d^4x d^4x'
     {
          [\partial_\mu A_\mu(x)] [\partial_\nu A_\nu(x')]
     \over
          (x-x')^2
     }~
$$
\beq
\label{4vect}
+~ surface~ terms
\eeq
\section{ Quantum Field Theory}

Returning now to quantum field theory and vacuum condensates, we
would like to examine the suggestion that \min , now the
expectation value of an operator, is sensitive to, or measures the
presence of topological structures in some way.

 A simple model we can investigate in this regard is 
 ``photodynamics'', i.e. the theory with the Lagrangian
density
\beq
L~=~ {1\over 4e^2}(F_{\mu\nu})^2\,.
\label{trivial}
\eeq
 This model can be studied in two realizations, 
compact and noncompact. While the noncompact
realization is just the theory of free photons, it is known
that the compact realization has nontrivial
properties, including a phase transition  near $e^2 \approx 1$ with
a condensation of magnetic monopoles \cite{polyakov}
(for review see, e.g. \cite{peskin}). Since the
monopoles are the sources of non-zero magnetic flux, we would
expect \min to be sensitive to the phase transition. 

\begin{figure}[h]
\centerline{
     \psfig{file=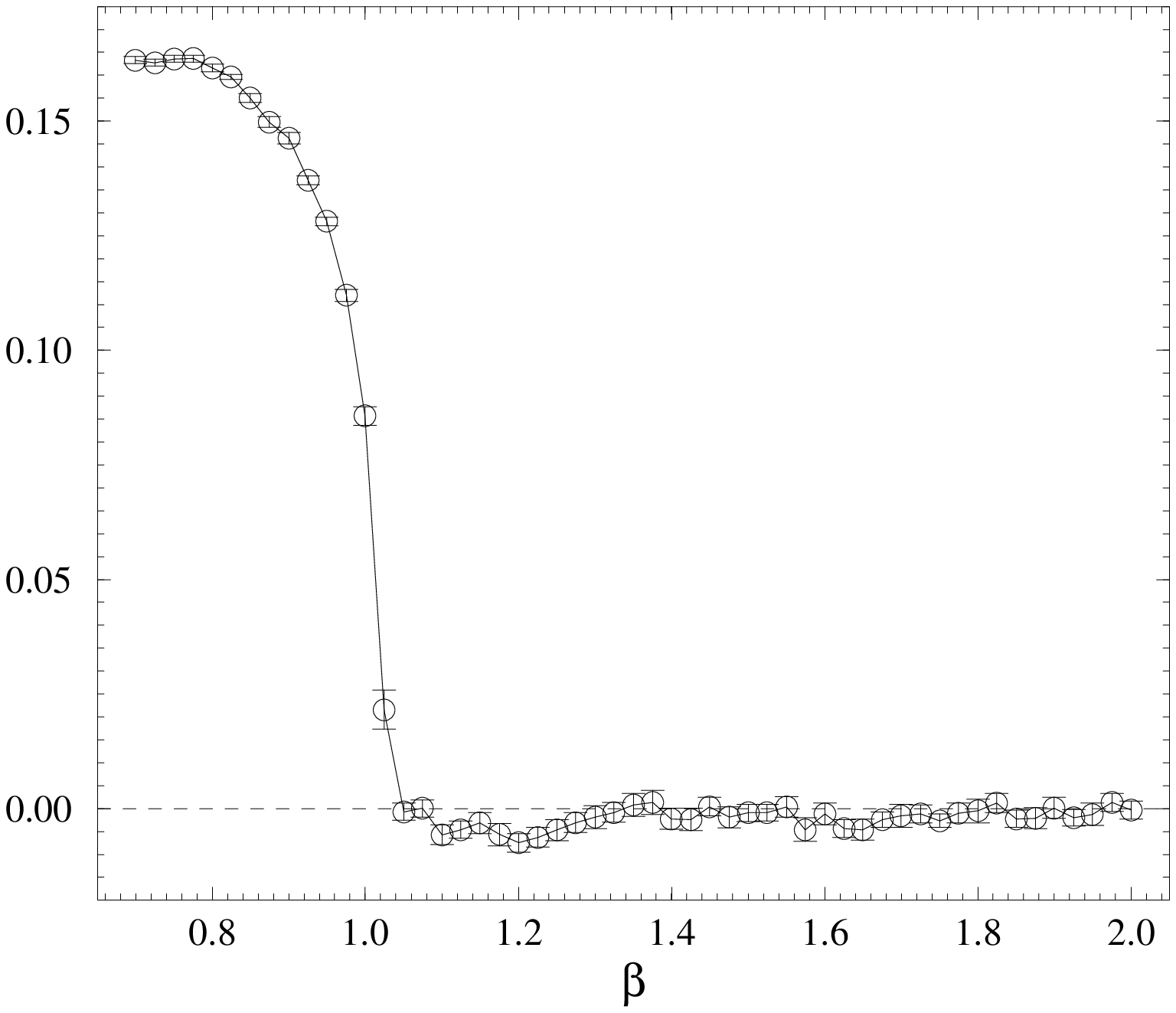,width=0.5\textwidth,silent=}
}
\noindent\fontsize{11}{13.2}\selectfont
Figure 1: $\zeta( e^2 )$, in units of the lattice spacing
as a function of $\beta = 1/e^2$ showing
the phase transition at $\beta = 1/e^2 \approx 1.0$.
\fontsize{12}{14.4}\selectfont
\end{figure}

 We can test these ideas in a numerical simulation by  considering 
the difference  of \min calculated in the two realizations. 
  We take the noncompact theory,  given by the  action (we
work in four euclidean dimensions)
\beq \label {non}
S_{non}(F)~=~ {1\over 4 e^2} \int d^4 x ( F_{\mu\nu} )^2\,,
\eeq
and the compact theory where
\beq \label {com}
S_{com}(F)~=~ {1\over 2 e^2} \int d^4 x [1- cos( F_{\mu\nu} )^2]\,,
\eeq
 and we would like to examine the difference 

\beq
\label{avg}
\zeta( e^2 )= \int\cD A~  A^2 e^{- S_{com}}- \int\cD A~  A^2
e^{- S_{non}} \,.
\eeq
  where $\int A^2$ is at its minimum for each
   gauge equivalent configuration.

 We do this in a lattice formulation, using a $12^4$ lattice with
periodic boundary conditions. $A^2$ is then measured in units of
the
lattice spacing. $\cD A~ $ is normalized so that $\int\cD A~e^{-
S}=1$, with $A(x)$ running  essentially from $-\infty$ to $+\infty$
in the noncompact case and from $-\pi$ to $+\pi$ in the compact
case. The ``minimum $A^2$" condition is
enforced by an iterative procedure: given a certain $A$
configuration on the lattice links, a gauge function $\alpha(x)$
(giving a new potential, $A-\nabla \alpha$) is repeatedly adjusted
so as to reduce the volume integral of $ A^2$ . Each pass works 
outward from an
arbitrary lattice point  and the procedure stops when the
reduction is less than a certain amount.

 Fig.~1 shows the results of the  numerical simulation. The sharp
jump in $\zeta( e^2 )$ at the phase transition supports the idea
that \min is a measure of the presence of the monoples and their
associated strings, present for $e^2\geq 1$.
The fact that $\zeta$ jumps to zero is related to the particularly
simple aspect of this model, that the small $e^2$ sectors of the
compact and noncompact theory have the same behavior.

With these numerical calculations we have studied the ground
state. When one  inserts an external monopole, it is also possible
to show the differing response of \min in the two theories by
analytic arguments \cite{gub}.

Many open and interesting questions remain, particularly
concerning the nonabelian case and the role of a d=2 condensate in
QCD. We hope to deal with some of them  in future work~\cite{gub}.


\section*{Acknowledgments}

We are thankful to  V.A.~Novikov and M.I.~Polikarpov for useful
discussions.



\begin{thebibliography}{72}

\bibitem{svz}
M.A.~Shifman, A.I.~Vainshtein, V.I.~Zakharov, \NP{B147}{1979}{385,
448}.

\bibitem{comment}
 An $\langle A^2 \rangle$ condensate in QCD was considered in the
past (by
M.J.~Lavelle, M.~Schaden, \PL{B 208}{1988}{297}, M.~Lavelle,
M.~Oleszczuk, \MPL{A 7}{1992}{3617} and references therein) 
in connection with gauge variant quantities like the gluon
propagator, and taken to be only of relevance if color conservation
is violated.


\bibitem{sc}
Our discussion also gives us an interesting insight into the
history of superconductivity, where
the Londons originally assumed that ${\bf A}$ was proportional to
the current, ${\bf A}\sim {\bf J}$.
This is a  striking statement where apparently a  gauge
non-invariant quantity is put equal to a gauge
invariant one. However note ${\bf A}\sim {\bf J}$ requires (for
static conditions)
${\bf \nabla A}= {\bf \nabla J}=0 $, which is the gauge
condition (\ref{diva}).
 This arises in the Ginzburg-Landau or Abelian Higgs
description for superconductivity:
in the  Free Energy or Langrangian there is the term
$|({\bf\nabla}  + e{\bf A})\psi|^2$ which  contains  $e{\bf A j}
+e^2 {\bf A}^2 |\psi|^2$, where
${\bf j}$ is the ordinary but  gauge non-invariant current $\psi^*
{\bf\nabla} \psi-\psi {\bf\nabla} \psi^*$,
while the full gauge invariant current is ${\bf J}={\bf j}-2e{\bf
A}|\psi|^2$. If we  write
$\psi= \sqrt\rho \, e^{i\phi}$, then ${\bf j}\sim \rho {\bf\nabla}
\phi$.  In the classical case of a
uniform superconductor with rigid superfluid density $\rho$,  we
can write ${\bf j} = {\bf\nabla}(\rho\phi)$. This is a pure
gradient
and  an integration by parts puts the
${\bf A j}$ term to  zero if
${ \bf\nabla} {\bf A}=0$. Only the ${\bf A}^2$ term in the
Lagrangian or Free Energy survives
and thus ${\bf A}$ interacts only with $ {\bf A}\rho$. (Note
however that  ${\bf J}$ itself, being a local
quantity, still contains ${\bf\nabla} \phi$). Hence  the Londons'
{\it Ansazt} may be
viewed as the  choice of  the ${ \bf \nabla} {\bf A}=0$ or
``minimum
$A^2$" gauge, together with the
physical input that $\rho$ is  non-zero and ``rigid".

\bibitem{kcross}
If ${\bf A}$ is a curl-free field this relation will be recognized
as
the expression for the electrostatic field energy in terms of the
charge density; if it is a divergence-free field, the magnetostatic
field energy in terms of the currents. The relation is essentially
the momentum space identity
$({\bf k} \times {\bf A})^2 = {\bf k}^2 A^2  -
({\bf kA})^2 $
in position space.
The  four dimensional version  is
$\frac{1}{2}(\epsilon_{\mu\nu\lambda\rho} k_{\lambda} A_{\rho})^2
=
k^2 A^2 - (k\cdot A)^2$.

\bibitem{polyakov}
A.M.~Polyakov, \PL{59B}{1975}{82}.

\bibitem{peskin}
M.E.~Peskin, \AP{113}{1978}{122}.

\bibitem{gub}
F.V.~Gubarev, V.I. Zakharov,
{\it ``On the emerging phenomenology of $\langle
(A^a_{\mu})^2_{min}
\rangle$"},
in preparation.


\end{thebibliography}
\end{document}